\documentclass[10pt,twocolumn,twoside]{IEEEtran}

\usepackage{graphicx}
\usepackage{subfigure}
\usepackage{setspace}
\usepackage{multicol}
\usepackage{multirow}
\usepackage{amsmath}
\usepackage{bm}
\usepackage{cite}
\usepackage{amssymb}
\usepackage{gensymb}
\usepackage{amsfonts}
\usepackage{mathrsfs}
\usepackage{amsmath}
\usepackage{algorithm}
\usepackage{algorithmic}
\usepackage{amsthm}

\usepackage{tabularx}
\usepackage{color}
\usepackage{balance}
\usepackage{mathrsfs}
\usepackage{setspace}
\usepackage{amsthm}
\usepackage{array}
\usepackage{cases} 
\newcommand{\PreserveBackslash}[1]{\let\temp=\\#1\let\\=\temp}
\newcolumntype{C}[1]{>{\PreserveBackslash\centering}p{#1}}
\usepackage[flushleft]{threeparttable}
\usepackage{setspace}
  
\usepackage{enumitem}

\begin{document}

\bibliographystyle{IEEEtran} 

\title{THz Precoding for 6G: Applications, Challenges, Solutions, and Opportunities}

\author{
	Jingbo Tan, \emph{Student Member, IEEE}, and Linglong Dai, \emph{Senior Member, IEEE}
	\vspace{-5mm}
	\thanks{All authors are with the Department of Electronic Engineering, Tsinghua University, Beijing 100084, P. R. China (E-mails: tanjb17@mails.tsinghua.edu.cn; daill@tsinghua.edu.cn).}

}

\maketitle
\IEEEpeerreviewmaketitle
\begin{abstract}
Benefiting from the ultra-wide bandwidth, terahertz (THz) communication is becoming a promising technology for future 6G networks. For THz communication, precoding is an essential technique to overcome the severe path loss of THz signals in order to support the desired coverage. In this article, we systematically investigate the dominant THz precoding techniques for future 6G networks, with the highlight on its key challenges and opportunities. Specifically, we first illustrate three typical THz application scenarios including indoor, mobile, and satellite communications. Then, the major differences between millimeter-wave and THz channels are explicitly clarified, based on which we reveal the key challenges of THz precoding, such as the distance-dependent path loss, the beam split effect, and the high power consumption. To address these challenges, three representative THz precoding techniques, i.e., analog beamforming, hybrid precoding, and delay-phase precoding, are extensively investigated in terms of their different structures, designs, most recent results, pros and cons. We also provide simulation results of spectrum and energy efficiencies to compare these typical THz precoding schemes to draw some insights for their applications in future 6G networks. Finally, several important open issues and the potential research opportunities, such as the use of reconfigurable intelligent surface (RIS) to solve the THz blockage problem, are pointed out and discussed.

\end{abstract}

\section{Introduction}
In February 2020, ITU officially launched the research work on 6G for the next-generation wireless networks. Various perspectives on 6G believe that 6G will empower full-dimensional wireless connectivity and coverage from indoor scenario to space scenario. Meanwhile, 6G will also support new vertical applications, such as holographic communications and extremely-high-definition video transmissions\cite{Ref:Survey6G2019}. To realize the 6G visions above, the peak data rate, which is an important key performance indicator for wireless communications, is expected to be more than 1 Tbps\cite{Ref:Survey6G2019,Ref:THzChannel2019}. However, the bandwidth provided by the 5G millimeter-wave band (e.g., 2 GHz bandwidth in 30-300 GHz\cite{Ref:EnHP2016}), is not able to support such a high data rate. Compared with mmWave band, terahertz (THz) band (0.1 THz-10 THz) can provide much more bandwidth, e.g., more than 20 GHz, to realize an ultra-high data rate\cite{Ref:THzChannel2019}. Thus, THz communication has been widely considered as an essential technology for future 6G networks\cite{Ref:THzChannel2019}. 

Nevertheless, it is known that THz signals seriously suffer from the severe path loss, e.g., 120 dB/100 m at 0.6 THz\cite{Ref:THzChannel2019}, which makes it difficult to realize the desired coverage. Precoding is one of the most promising solutions to solve this problem without increasing the transmit power. Specifically, by generating directional beams with high array gains using large-scale antenna arrays, precoding can concentrate the transmitted power in some specific directions to compensate for the severe path loss\cite{Ref:TeraSub2016}. Generally, the array gain of the generated beam is proportional to the scale of antenna array. Thanks to the very small wavelengths of THz signals, very-large-scale antenna arrays can be used in THz communications\cite{Ref:THzDis2016}. Consequently, THz precoding can generate pencil beams with very narrow beam width and very high array gains, which can significantly compensate for the severe path loss of THz signals. As a result, THz preoding is an indispensable technique for future 6G networks\cite{Ref:THzChannel2019}. Unfortunately, although precoding has been extensively studied in 5G mmWave systems, there are some new challenges for THz precoding due to the different characteristics of THz channels compared with mmWave channels, which should be identified and solved to enable a powerful 6G system.

In this article, we systematically investigate the dominant THz precoding techniques for future 6G networks, with the highlight on its key challenges and potential opportunities. Specifically, three typical THz application scenarios including the indoor, mobile, and space communications, are illustrated at first. Then, we clarify the major differences between mmWave and THz channels, based on which the key challenges for THz precoding are revealed, such as the distance-dependent path loss, the beam split effect, and the high power consumption. To address these challenges, we summarize the evolution of THz precoding technique, and highlight three typical THz precoding techniques, i.e., analog beamforming, hybrid precoding, and delay-phase precoding. Systematic comparison of these three THz precoding techniques is discussed in terms of their different structures, designs, most recent results, pros and cons. In addition, we also provide simulation results to compare these THz precoding schemes to draw some insights for their applications in future 6G networks. More importantly, in the end of this article, we point out several important open issues and discuss the potential research opportunities for THz precoding, such as the use of reconfigurable intelligent surface (RIS) to solve the blockage problem. We believe that this article would inspire and stimulate more innovative ideas and solutions for this important research topic of THz precoding for future 6G wireless networks. 

\section{THz Precoding: Applications and Challenges}
In this section, we first illustrate three typical THz application scenarios for future 6G networks. Then, we introduce the fundamental principle of precoding to compensate for the severe path loss of THz signals. Finally,  we identify the major differences between THz and mmWave channels, based on which the key challenges for THz precoding will be revealed.

\subsection{Typical THz Application Scenarios}
6G network will become an intelligent network that integrates Internet of things, indoor communications, mobile communications, underwater communications, and satellite communications\cite{Ref:Survey6G2019}. Here, we briefly illustrate three typical THz application scenarios for future 6G wireless networks.
\subsubsection{Indoor Communications}
Indoor scenario is the earliest research area of THz communications. By utilizing the ultra-wide bandwidth provided by THz band, the requirement of high-data-rate indoor transmission can be fulfilled. For example, IEEE 802.15.3d has already established the standard of THz physical layer in the band of 0.252-0.325 THz to support short-distance transmission with 100 Gbps data rate\cite{Ref:80215Stan2017}. 
\subsubsection{Mobile Communications}
To support higher data rate transmission for mobile users,  THz communications are expected to be adopted for hot spot areas within a micro-cell. In addition, THz communications can also support ultra speed backhaul data transmission between base stations \cite{Ref:Survey6G2019}. 

\subsubsection{Satellite Communications}
To realize full-dimensional coverage for future 6G network, satellite communications is an indispensable scenario\cite{Ref:Survey6G2019}. On one hand, with the help of THz communications, the satellite-terrestrial integrated networks can provide high-data-rate connectivity for people living in remote areas without sufficient information infrastructure\cite{Ref:BeamformingSate2019}. On the other hand, THz communications can enable high-data-rate space links between satellites or space vehicles, which is critical for space exploration.

\subsection{Principle of THz Precoding}
Among the above application scenarios of THz communications, a common important issue is how to deal with the severe path loss of THz signals. THz precoding is an essential technique to solve this problem. In this subsection, we will explain the fundamental principle of THz precoding. 

Precoding is a channel adaptive technique, which pre-processes  the transmitted signal based on the channel information at the transmitter\cite{Ref:EnHP2016}. The most important function of THz precoding is to generate directional pencil beams to compensate for the severe path loss of THz signals, and it can also mitigate the multi-user interferences. The basic principle to generate a narrow pencil beam in a certain direction is to make electromagnetic waves that transmitted by different antenna elements form an equiphase surface, which is perpendicular to the desired propagation direction\cite{Ref:EnHP2016}. To realize this goal, incremental phase shifts that increases by the label of antenna element  should be compensated at different antenna elements. Generally, the power gain obtained by the generated beam is called as array gain, which is proportional to the antenna number, and the width of the beam is inversely proportional to the antenna number. Considering the antenna spacing is usually set as half of the wavelength and the wavelength of THz signals is very small, we can employ very-large-scale antenna array (e.g., 1024-element\cite{Ref:THzChannel2019}) in THz communications, which is much larger than that in mmWave communications (e.g., 256-element\cite{Ref:EnHP2016}). Therefore, the severe path loss can be significantly compensated by a high-array-gain pencil beam, e.g., 60 dB array gain\cite{Ref:THzChannel2019}.

\subsection{Key Challenges}\label{CH}

\begin{table*}
	\caption{Comparison Between mmWave and THz Channels}\label{Table1}
	\centering
	\begin{tabular*}{0.76\textwidth}{c|c|c}
		\hline
		& mmWave & THz  \\
		\hline
		Frequency band& 30-300 GHz& 0.1-1 THz\\
		\hline
		Bandwidth& Wide bandwidth & Ultra-wide bandwidth\\		
		\hline
		Molecular absorption & Not significant & Significant\\ 
		\hline
		Propagation path number& A few paths & Line-of-sight (LoS) path dominant\\ 
		\hline
		Propagation path loss& Severe loss & More severe and distance-dependent\\
		\hline
		Number of antennas& large-scale antenna array &
		Very-large-scale antenna array\\	
		\hline
		Beam width& Narrow beam &
		 Pencil beam\\
		 \hline
		Wideband effect & Beam squint & Beam split\\
		\hline
	\end{tabular*}
\end{table*}

As one of the key techniques, precoding has already been extensively investigated in 5G mmWave networks. Nevertheless, due to different characteristics of THz channels compared with mmWave channels, there are some new challenges for THz precoding. Table I shows the major differences between mmWave and THz channels\cite{Ref:THzChannel2019}, based on which the key challenges for THz precoding are explained as follows.

\begin{figure}
	\centering
	\includegraphics[width=0.47\textwidth]{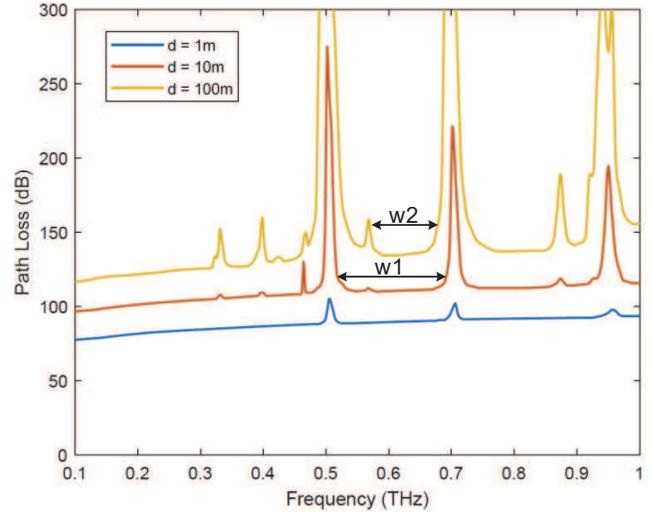}
	\caption{Distance-dependent path loss in THz band.}
\end{figure}

\subsubsection{Distance-Dependent Path Loss}
Due to the significant molecular absorption, the propagation distance of THz signal will not only determine the value of path loss, but also affect the available bandwidth\cite{Ref:TeraNets2014}. Specifically, Fig. 1 illustrates the path loss in the THz band against the frequency for different transmission distances\cite{Ref:TeraNets2014}, where the sharp increase of path loss is induced by the molecular absorption at some frequencies. We can observe that the available bandwidth window is distance-dependent in THz band, e.g., the available bandwidth window w1 for the transmission distance of 10 m is 0.51-0.68 THz, while the available bandwidth window w2 for the transmission distance of 100 m is 0.56 -0.67 THz. This distance-dependent path loss caused by the molecular absorption is not significant in mmWave band, but quite obvious in THz band. Since most existing precoding methods are mainly using unified available bandwidth window to serve different users at different distances, they may suffer from the severe performance loss in THz communications. Therefore, the distance-dependent path loss has to be addressed for THz precoding\cite{Ref:THzDis2016}.

\subsubsection{Beam Split Effect}
To carry out precoding, the most widely utilized signal processing components to realize phase shifts are analog phase-shifters (PSs). Generally, the required phase shifts are related to the wavelength, which is determined by the carrier frequency. This is not a matter for narrowband systems with a small frequency range. However, in wideband systems with a large frequency range, since the PSs can only realize \emph{frequency-independent} phase shifts, the beams generated by PSs may disperse to the surrounding directions, and thus introduces the array gain loss\cite{Ref:DPP2019}. This effect is called as beam squint in mmWave systems\cite{Ref:EnHP2016}, where the array gain loss is not serious. However, as shown in Fig. 2, the ultra-wide bandwidth and the very narrow pencil beam in THz systems will significantly aggravate this effect, where the beams at different frequencies may split in different directions far away from the target user. Therefore, this beam split effect will result in much more severe array gain loss for THz precoding, and consequently lead to the unacceptable degradation of the achievable rate performance\cite{Ref:DPP2019}.

\subsubsection{High Power Consumption}
With the employment of very-large-scale antenna array, the huge power consumption is an inevitable problem for THz precoding. Communication systems with multiple antennas in low-frequency-band usually utilize the fully-digital precoding, where each antenna element is connected to a dedicated radio-frequency (RF) chain, so the phase shifts can be easily realized by baseband signal processing. However, since a single RF chain in THz band requires large power consumption (e.g., 250 mW\cite{Ref:PSorSW2016}), a large number of antennas will result in huge power consumption. On the other hand, although the power consumption can be relieved by replacing RF chains with analog components such as PSs (e.g., 30 mW\cite{Ref:PSorSW2016}), the large number of PSs resulted by the large number of antennas makes THz systems still suffer from high power consumption. Therefore, a low-power hardware structure is vital for THz precoding.

\section{THz Precoding Techniques}
To relieve the hugh power consumption, THz precoding techniques are mainly based on low RF-complexity solutions by introducing analog components\cite{Ref:THzChannel2019}. In this section, we will highlight three representative THz precoding techniques with low RF-complexity as shown in Fig. 3, i.e., analog beamforming, hybrid precoding, and delay-phase precoding. 

\begin{figure}
	\centering
	\includegraphics[width=0.47\textwidth]{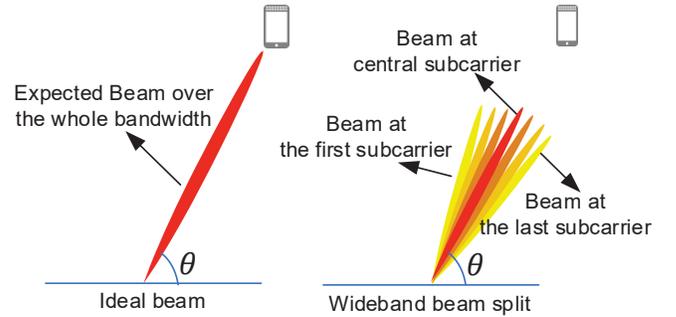}
	\caption{THz wideband beam split effect.}
\end{figure}

\subsection{Analog Beamforming}
\subsubsection{Hardware Structure}
Analog beamforming is a basic precoding technique. The key idea is to utilize only one RF chain and a PSs network to reduce RF-complexity\cite{Ref:EnHP2016}. The hardware structure of analog beamforming is shown in Fig. 3, where the RF chain is connected to all antenna elements via PSs.
\subsubsection{Analog Beamforming Design}
Since only one RF chain is employed in analog beamforming, only one user with single transmitted data stream can be served. Under this circumstance, the optimal beamforming design is to generate a beam towards the physical direction with the highest channel power, which is usually the LoS path direction in THz channel. In this way, the optimal array gain can be achieved.

\subsubsection{Pros and Cons}
By utilizing analog PSs and only one RF chain, the analog beamforming requires quite low power consumption, where the number of PSs is equal to the antenna number. However, analog beamforming can only support single-stream transmission, which cannot be easily applied in multi-user or multi-stream scenarios\cite{Ref:EnHP2016}. Besides, analog beamforming can only adjust the phase of beamforming vector due to the hardware constraint of PSs, which will result in some performance loss\cite{Ref:EnHP2016}. The features of single-stream transmission and low power consumption make analog beamforming suitable for long-distance single-stream transmission.

\begin{figure*}
	\centering
	\includegraphics[width=0.88\textwidth]{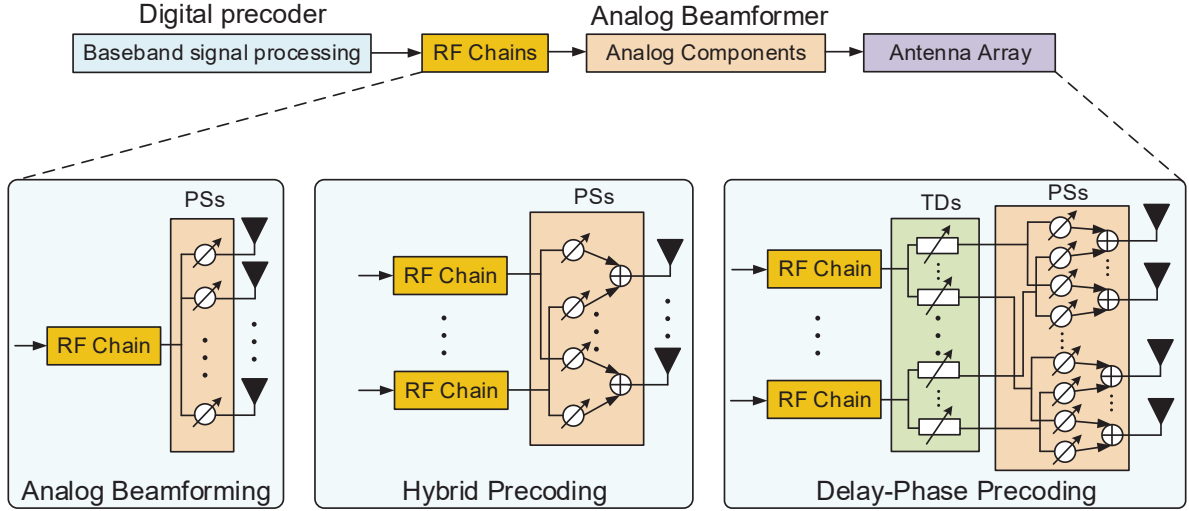}
	\caption{THz precoding techniques: analog beamforming, hybrid precoding, and delay-phase precoding.}
\end{figure*}

\subsection{Hybrid Precoding}
\subsubsection{Hardware Structure}
To solve the problem that the analog beamforming with one RF chain can only support single-stream transmission, hybrid precoding with a few RF chains has been proposed, as shown in Fig. 3\cite{Ref:Twostage2015,Ref:EnHP2016}. The key idea of hybrid precoding is to realize precoding by a large-dimensional analog beamformer realized by analog components (e.g., PSs) and a small-dimensional digital precoder realized by a small number of RF chains, where the number of RF chains is much smaller than the antenna number (e.g., 4 RF chains for 1024 antenna elements). Therefore, multiple streams can be transmitted simultaneously. There are several connection structures between RF chains and PSs. For example, the full-connected structure is shown in Fig. 3\cite{Ref:Twostage2015}, where each RF chain is connected to all antenna elements via PSs, so the analog beamformer is a full matrix. Except for the full-connected structure, the sub-connected structure is also widely considered\cite{Ref:EnHP2016}, where each RF chain is only connected to a subset of antenna elements and the antenna elements connected by different RF chains are non-overlapping, so the analog beamformer becomes a block diagonal matrix in this case. 

\subsubsection{Hybrid Precoding Design}The precoding design problem is to design analog beamformer and digital precoder based on channel information to achieve the optimal achievable sum-rate performance. The hybrid precoding design can be divided into two categories: two-stage precoding design and joint optimization based precoding design.
\begin{itemize}[leftmargin=3.7mm]
\item{\textbf{Two-stage precoding design:} The optimization problem of hybrid precoding design is difficult to solve, due to the non-convex hardware constraint of analog beamformer. A representative near-optimal method is two-stage precoding, where the analog beamformer and  digital precoder are designed separately. Specifically, for the full-connected structure\cite{Ref:Twostage2015}, the optimal analog beamforming is carried out for each user to maximize the array gain in the first stage. Then, in the second stage, the traditional zero-forcing digital precoder is utilized to cancel the inter-user interferences. A similar two-stage precoding method for sub-connected structure was proposed in \cite{Ref:TeraSub2016}. In the first stage, users with close physical directions are selected as a group, and they are served by an analog beam generated by several subsets of antenna elements. Then, in the second stage, by considering the distance-dependent path loss, the frequnecy band for each user is properly allocated, and zero-forcing based digital precoder is designed to optimize the sum-rate performance, which mitigates the performance loss induced by the distance-dependent path loss of THz signals.}

\item{\textbf{Joint optimization based precoding design:} Another category to design hybrid precoder is to jointly optimize the analog beamformer and digital precoder. Specifically, \cite{Ref:AltMinPre2016} proposed an alternately minimization based method for the full-connected structure, which alternately optimizes the digital precoder and analog beamformer without any restriction on the analog beamformer. For the sub-connected structure, a successive interference cancellation (SIC) based method was proposed in \cite{Ref:EnHP2016}. It firstly decomposes the non-convex hybrid precoding design problem into several convex sub-problems, each of which optimizes a subset of antenna elements. Then, the hybrid beamformer is obtained one-by-one following the procedure in SIC-based signal detection. Note that the complexity of joint optimization methods is usually higher than that of the two-stage methods.}
\end{itemize}

As revealed in Section \ref{CH}, it is a crucial challenge to mitigate the beam split effect for THz precoding. However, in wideband systems, the two-stage precoding designs based on \emph{frequency-independent} PSs \cite{Ref:Twostage2015} will suffer from severe performance loss, since analog beams cannot be aligned with users. On the other hand, the joint optimization based precoding design can relieve the performance loss by directly optimizing the achievable rate\cite{Ref:AltMinPre2016}, but the performance is still restricted by PSs. Actually, the key idea to eliminate beam split effect is to introduce \emph{frequency-dependent} components. Following this idea, an effective solution is to replace all PSs by time-delayers \cite{Ref:TeraSub2016}. Utilizing the frequency-dependent phase shifts provided by time-delayers, the beams can be aligned with users over the whole THz bandwidth, and thus the severe performance loss caused by the beam split effect can be mitigated.

\subsubsection{Pros and Cons}
Compared with analog beamforming, hybrid precoding is able to achieve a better trade-off between achievable rate performance and power consumption. Multi-user or multi-stream transmissions can be supported. Specifically, the full-connected hybrid precoding structure can generate multiple beams, the number of which is equal to the number of RF chains. Since each beam utilizes all antenna elements, the optimal array gain and narrow beam width can be expected, which leads to the satisfying performance. However, the number of PSs in the full-connected structure that is equal to the number of RF chains times the antenna number is quite large, which will still cause huge power consumption. In contrast, the sub-connected structure has the same number of PSs and antenna elements, which can further reduce the power consumption. Nevertheless, since only part of antenna elements (a subset or several subsets) are exploited to generate a beam for a certain user, the sub-connected structure usually suffers from the array gain loss. 

\vspace{-2mm}
\subsection{Delay-Phase Precoding}
\subsubsection{Hardware Structure}
As we have mentioned in the subsection III-B above, the wideband beam split effect is a key challenge for THz precoding, which can be addressed by replacing traditional PSs with time-delayers. In this way,  a large number of time-delayers are required for the full-connected structure. However, THz time-delayers consumes much higher power (e.g., 80 mW\cite{Ref:TTDpower2016}) than traditional PSs (e.g., 30 mW\cite{Ref:PSorSW2016}). Therefore, the large number of time-delayers significantly increases the power consumption. To compensate for the beam split effect with low power consumption, a delay-phase precoding structure has been recently proposed in \cite{Ref:DPP2019}. As shown in Fig. 3, it introduces a time-delayers network between RF chains and traditional PSs network to realize \emph{frequency-dependent} analog beamforming, but it has a significantly reduced number of time-delayers. Specifically, each RF chain is connected to several time-delayers, where the number of time-delayers is much smaller than the antenna number (e.g., 32 time-delayers per RF chain for 1024 antenna elements), and these time-delayers are connected to all the antenna elements via PSs in a sub-connected manner.

\subsubsection{Delay-Phase Precoding Design}
The signal model of delay-phase beamforming is essentially the same as that of hybrid precoding, while the key difference lies in the analog beamformer design. Specifically, unlike hybrid precoding where the analog beamformer is solely realized by PSs, the analog beamformer in the delay-phase precoding is divided into two concatenated parts. One is realized by PSs, and the other is realized by time-delayers. \cite{Ref:DPP2019} has proposed a two-stage precoding design method for delay-phase precoding. In the first stage, the PSs generate frequency-independent beams towards different users similar to that in \cite{Ref:Twostage2015}. Then, the time delays provided by time-delayers connected to a certain RF chain is designed by a beam direction compensation mechanism based on the user physical direction and THz signal bandwidth. Therefore, the frequency-dependent beams can be aligned with different users over the whole ultra-wide bandwidth. In the second stage, zero-forcing precoding is carried out separately for different subcarriers to mitigate multi-user interferences.

\subsubsection{Pros and Cons}
Delay-phase precoding converts the traditional one-dimensional analog beamformer into two-dimensional analog beamformer, i.e., the sole control of the phase shifts is extended to the joint control of phase shifts and time delays. It has been proved that delay-phase precoding can achieve the near-optimal array gain over the whole bandwidth, while the required number of time-delayers can be significantly reduced (e.g., for 1024 antenna elements with 4 RF chains, the number of time-delayers can be reduced from 4096 to 128\cite{Ref:DPP2019}). Hence, the power consumption can be significantly reduced.

\vspace{-2mm}
\section{Performance Comparison}
In this section, we provide simulation results to compare the achievable sum-rate and energy efficiency of different precoding techniques, together with the classical fully-digital precoding\cite{Ref:EnHP2016}. We consider a single-cell multi-user THz system with the bandwidth of 20 GHz at carrier frequency 350 GHz, where the base station equips a 1024-element uniform linear array to serve several single-antenna users. We consider Saleh-Valenzuela channel model \cite{Ref:AltMinPre2016} with a single LoS path. The performance of the full-connected and sub-connected hybrid precoding with PSs and time-delayers (TDs), and the delay-phase precoding structure are compared, where the number of time-delayers is 32 and the number of RF chains is equal to the user number.  Two-stage precoding methods\cite{Ref:Twostage2015,Ref:TeraSub2016,Ref:DPP2019} are carried out.
\begin{figure}
	\centering
	\includegraphics[width=0.45\textwidth]{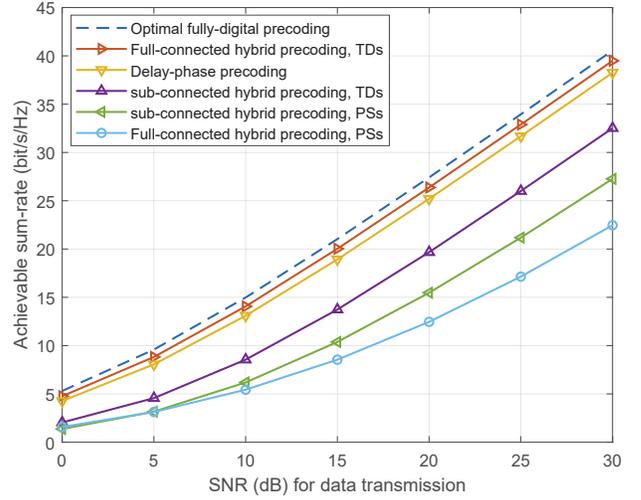}
	\caption{Achievable sum-rate against SNR.}
\end{figure}

Fig. 4 shows the achievable sum-rate performance, where we set the number of users as 4. We can observe from Fig. 4 that the hybrid precoding structures with PSs suffer from severe achievable sum-rate loss compared with the optimal fully-digital precoding. This is because they cannot deal with the beam split effect. On the contrary, with the help of TDs, hybrid precoding structure using TDs can achieve better performance. Moreover, although the number of TDs is quite small, the delay-phase precoding can also achieve the near-optimal performance, which outperforms the sub-connected hybrid precoding with TDs by about 5 dB.

Fig. 5 illustrates the energy efficiency against the number of users. We define the energy efficiency as the ratio between the achievable sum-rate and the summation of the transmitted power and the hardware power. Specifically, the transmitted power is set as 2.5 W\cite{Ref:EnHP2016}. The hardware consumption is computed based on the structures shown in Fig. 3, where the power of baseband processing, RF chain, PS, and TD are set as 250 mW, 250 mW, 30 mW, and 80 mW\cite{Ref:PSorSW2016,Ref:TTDpower2016}, respectively. We can observe that the full-connected structure achieves poor energy efficiency because of the huge number of PSs or TDs. The sub-connected structure enjoys higher energy efficiency when the number of users is larger than 4. This is because the fact that the number of PSs or TDs in sub-connected structure is fixed as the antenna number. When the number of users is less than or equal to 4, the delay-phase precoding enjoys the highest energy efficiency. This is because the achievable sum-rate loss caused by the beam split effect can be mitigated with a relatively small number of TDs when the number of users is small.
\begin{figure}
	\centering
	\includegraphics[width=0.45\textwidth]{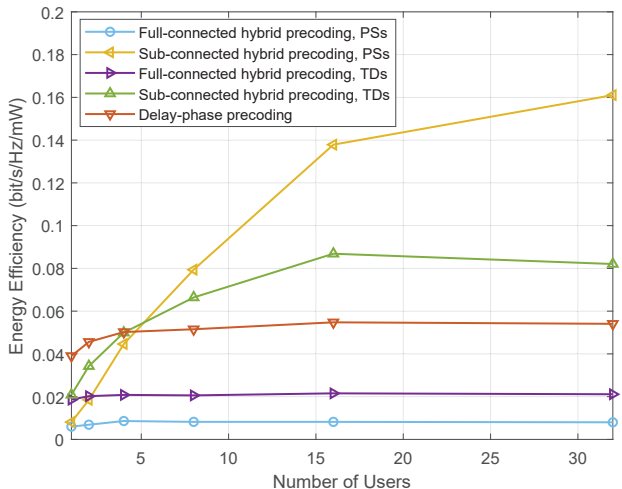}
	\caption{Energy efficiency against the number of users.}
\end{figure}

\section{Open Issues and Research Opportunities}
In this section, we point out several open issues for THz precoding to empower future 6G networks, and the corresponding research opportunities are also discussed.

\subsection{Blockage}
THz communication mainly relies on the LoS path. When the LoS path is blocked, severe performance loss will occur. To solve this blockage problem, the recently proposed RIS can be utilized. Specifically, by using RIS as a passive relay between the BS and users, RIS can establish a new reliable path other than the LoS path, and realize flexible beamforming by adjusting the phases of many RIS elements\cite{Ref:RIS2020}. In this way, the blockage of THz signals can be solved. For RIS-aided THz precoding, the performance analysis and joint precoding design are still not clear, which requires further study in the near future.

\subsection{Hardware Impairments}

Existing THz precoding studies are mainly based on the ideal hardware assumption. However, the actual performance of THz precoding will be degraded by inevitable hardware impairments\cite{Ref:THzChannel2019}, such as the non-linearity of power amplifier, In-phase/Quadrature imbalance, phase noise, and so on. Therefore, for THz wideband system, accurate THz hardware impairments model, exact performance analysis, and the corresponding beamforming calibration designs are still open problems to be investigated.

\subsection{Lower Power Consumption}

Although the three representative THz precoding techniques discussed in this article can relieve the huge power consumption, they still consume a large amount of power. Therefore, further reduction of the power consumption is expected for the practical implementation of THz systems, such as using low-resolution ADC/DAC or low-resolution PSs. Particularly, the recently proposed RIS can replace the existing phased array to significantly reduce the power consumption\cite{Ref:RIS2020}. The THz precoding design based on these low-power analog components has not been addressed well in the literature, which requires further investigation from both hardware design and signal processing perspectives.

\subsection{User Mobility}
Due to the very narrow pencil beams in THz systems, the optimal precoder will vary fast when users are moving. Therefore, THz precoding considering user mobility is an important issue. For instance, efficient beam tracking or channel tracking schemes can quickly calibrate the beam direction and guarantee the continuous service. However, the existing precoding methods considering user mobility are designed for mmWave channels, which cannot deal with the new challenges in THz communication systems. Hence, THz precoding considering user mobility should be addressed in the future.

\subsection{Scintillation Effect}
In THz band, the temperature and humidity of environment can aggravate the turbulence of THz signals, which causes the inevitable scintillation effect\cite{Ref:SiEffect2019}. The scintillation effect will impose a fast-varying non-linear disturbance on the THz channel, which induces a severe performance degradation. However, there is very little research on THz communications considering the scintillation effect. The signal model of THz scintillation effect and corresponding signal processing schemes requires further research, where the machine learning algorithms good at solving non-linear problems may be helpful.


\subsection{Other issues}
There are some other open issues for THz precoding, such as the accurate THz channel model, the complex design and fabrication of very large THz antenna array, the low-power design of THz analog components, etc., which require further investigation in the near future.

\section{Conclusions}
In this article, we systematically investigated the dominant THz precoding techniques for future 6G networks. Based on the major differences between mmWave and THz channels, we revealed the key challenges for THz precoding, and compared three representative THz precoding techniques. Specifically, analog beamforming has the simplest hardware structure, but can only support single-stream transmission. Hybrid precoding achieves a better tradeoff between performance and power consumption. Delay-phase beamforming can solve the performance loss caused by the beam split effect with much lower power consumption.  Finally, we identified several research opportunities for THz precoding to empower future 6G networks.

\bibliography{tjb2}
\section*{Acknowledgment}
This work was supported by the National Science and Technology Major Project of China under Grant 2018ZX03001004-003 and the National Natural Science Foundation of China for Outstanding Young Scholars under Grant 61722109.

\section*{Biographies}

\textbf{Jingbo Tan} [S'17] received his B. S. degree in the Department of Electronic Engineering,  Tsinghua University, Beijing, China, in 2017, where he is currently pursuing his Ph. D. degree. His research interests are mmWave and THz communications.
\\

\textbf{Linglong Dai} [M'11, SM'14] received the Ph.D. degree from Tsinghua University, Beijing, China, in 2011. He is currently an associate professor at Tsinghua University. His current research interests include RIS, massive MIMO, mmWave/THz communications, and machine learning. He has received six conference best paper awards and four journal best paper awards.

\end{document}